\documentclass[12pt,a4paper]{article}
\begin{document}
\textwidth=135mm
 \textheight=200mm
\begin{center}
{\bfseries
Peculiar features of the
relations  between pole and running heavy quark masses
and    estimates  of the  $O(\alpha_s^4)$ contributions
}
\vskip 5mm
A. L. Kataev$^\dag$  and V.
T. Kim$^\ddag$
\vskip 5mm
{\small {\it $^\dag$ Institute for
Nuclear Research of the Academy of Sciences of Russia,
 117312  Moscow, Russia}} \\
{\small {\it $^\ddag$ St. Petersburg Nuclear Physics Institute of Russian Academy
of Sciences 188300, Gatchina, Russia}}
\\
\end{center}
\vskip 5mm
\centerline{\bf Abstract}
Perturbative  relations between pole and running heavy quark masses,
defined in the Minkowski
regions,  are considered.
Special attention is paid to the appearance  of
the kinematic $\pi^2$-effects, which exist in the coefficients
of these series.
The estimates of order $O(\alpha_s^4)$ QCD  corrections
are presented.
\vskip 10mm
\section{Introduction}
Among important parameters of QCD are the
masses of  $c$, $b$ and $t$-quarks,
which  are more heavy than $N_L$=3,4,5
number of lighter ones. They can be defined either
as the poles of the corresponding renormalised
heavy-quark propagators at $q^2=M_{(N_L+1)}^2$
 in the Minkowski space-like region
or as the running masses
$\overline{m}_{(N_L+1)}(\mu^2)$ in the  $\overline{\rm MS}$-scheme.
Their scale-dependence is described by the solution of the
following equation
\begin{equation}
\label{RG}
\frac{\overline{m}_{(N_L+1)}(s)}{\overline{m}_{(N_L+1)}(\mu^2)}
= {\rm exp} \bigg[\int_{a_s(\mu^2)}^{a_s(s)} \frac{\gamma_{m_{(N_L+1)}}(x)}
{\beta(x)}dx \bigg]
\end{equation}
where $a_s(s)=\alpha_s(s)/\pi$ and $\alpha_s(s)$ is the QCD coupling
constant of the  $\overline{\rm{MS}}$-scheme, fixed in the {\bf Minkowski}
reference point  $s>\overline{m}_{(N_L+1)}^2$,  and
the renormalization group  functions $\gamma_{m_{(N_L+1)}}(x)$
and $\beta(x)$ are defined as
\begin{eqnarray}
\label{gamma}
\gamma_{m_{(N_L+1)}}(a_s)&=&
\frac{d\ln{\overline{m}_{(N_L+1)}}(\mu^2)}{d \ln \mu^2}
= - \sum_{i\geq 0 }  \gamma_i(N_L) a_s^{i+1}~~\\
\label{beta}
\beta(a_s)&=&\frac{d a_s(\mu^2)}{d \ln \mu^2}=
-\sum_{i\geq 0}\beta_i(N_L) a_s^{i+2} .
\end{eqnarray}
The coefficients  $\beta_i(N_L)$ and $\gamma_i(N_L)$
(apart of the coefficient $\gamma_0$) depend from $N_L+1$ number
 of active flavours. Note, that  for the   $\overline{\rm MS}$-scheme
heavy quarks running masses  $\overline{m}_{(N_L+1)}(\mu^2)$ the {\bf
Minkowskian} normalization point $\mu^2=\overline{m}_{(N_L+1)}^2$
is frequently used  (see, e.g.,
Ref. \cite{Chetyrkin:2000yt}).    In this case the definition of
$\overline{m}_{(N_L+1)}(\overline{m}_{(N_L+1)}^2)$
may be  geometrically
illustrated  by finding the intersection of the curve, which represent the
inverse logarithmic
scale-dependence of the squared running mass,  with the bisectrix
of the  angle, formed by   positive axises
$0\leq \overline{m}_{(N_L+1)}^2 \leq \infty$ and  $0 \leq \mu^2 \leq \infty$
\footnote{We are grateful to G.B. Pivovarov for the discussion of this topic.}.
The relations  between pole and running heavy quark masses
we will be
interested read
\begin{equation}
\label{massrel}
M_{(N_L+1)}=\overline{m}_{(N_L+1)}(\overline{m}_{(N_L+1)}^2)\sum_{n=0}^{4}
 t_n^{M}(N_L) a_s^n(\overline{m}^2_{(N_L+1)})~~.
\end{equation}
Note, that in the process of comparison of  theoretical predictions
for the $e^+e^-$-annihilation Euclidean
time-like characteristic, namely
Adler D-function,  with its
experimental-motivated behaviour  \cite{Eidelman:1998vc}
heavy  quark pole masses were defined in the MOM-scheme, while
heavy quark running  masses were
defined at  the ${\bf Euclidean}$ scale   $\mu^2=Q^2$.
The similar mixed MOM- $\overline{\rm{MS}}$-scheme prescriptions
are  also  widely used to analyse  heavy-quark mass dependent effects
in characteristics   of deep inelastic scattering
(see e.g. \cite{Bierenbaum:2009mv}, \cite{Forte:2010ta}). However,
the processes, which may be observed at LHC, are described by
theoretical
predictions in the  time-like region of energies.
In view of this it is important to study
relations between different most commonly used  definitions
of heavy quark masses and to derive the relations between pole
and running heavy quark masses, tied   to  the {\bf Euclidean} and
{\bf Minkowski} regions of momentum transfered.
This problem was   analysed in Ref.\cite{Chetyrkin:1997wm}
with the help of  the special   K\"allen-Lehman
type representation. Here we will consider this approach
in more detail, presenting additional arguments in favour of
theoretical background of the investigations, performed  in the work
mentioned above.
We  will also  update estimates of
the order $O(\alpha_s^4)$ terms in
the relation of Eq.(\ref{massrel}), which were  obtained in Ref.\cite{Chetyrkin:1997wm}
using the  extended to the  mass-dependent case effective-charges inspired
massless approach, elaborated in Ref.\cite{Kataev:1995vh}.

\section{Comments on application of the dispersion relations}
Let us discuss the subject of applicability  of  the K\"allen-Lehman
 type spectral representations  within the
context of perturbative QCD.
The well-defined dispersion relation  for  the
$e^+e^-$-annihilation Adler function
is well known
\begin{equation}
\label{DA}
D_V(Q^2)=-Q^2\frac{d\Pi_V(Q^2)}{dQ^2}=Q^2\int_0^{\infty}\frac{R(s)}{(s+Q^2)^2}ds
\end{equation}
where $\Pi_V(Q^2)$ is the photon vacuum polarization function and
$R(s)\sim Im \Pi_V $.
The two-point function
of the scalar quark currents $m_{(N_L+1)}{\overline{\psi}}_q{\psi}_q$
has  the imaginary part, which  defines
the scalar Higgs
boson decay width into quark-antiquark pairs. In this case
it is possible to write-down the following representation
\cite{Gorishnii:1983cu}:
\begin{equation}
\label{DS}
D_S(Q^2)=-Q^2\frac{d}{dQ^2}\bigg[\frac{\Pi(Q^2)}{Q^2}\bigg]=
Q^2\int_0^\infty\frac{R_S(s)}{(s+Q^2)^2}ds~~
\end{equation}
 which faces no problems in the region where the the asymptotic freedom
property of QCD holds.  The same equation
was used in Ref.\cite{Chetyrkin:1997wm} to extend the
massless procedure
of the estimates of higher-order perturbative   corrections to the
Euclidean quantities \cite{Kataev:1995vh}
to the case of Eq.(\ref{DS}),
which contains the dependence from the square of running mass
$\overline{m}_{(N_L+1)}(Q^2)$ defined in the {\bf Euclidean} region.
However, as was
shown in Ref.\cite{Broadhurst:2000yc}, the dispersive relation of Eq.(\ref{DS})
is valid within perturbative sector only and can not be proved
 on the level of rigour, considered in
Ref.\cite{Bogolyubov-Medvedev-Polivanov}. Indeed, it was shown in Ref.
\cite{Broadhurst:2000yc} that in the low-energy region
Eq.(\ref{DS}) is ill-defined and contains fictitious
$\Lambda_{QCD}^2/Q^2$-term. It
reflects  the failure to remove the infinities from $\Pi_S(0)$.
The well-defined dispersive relation,  which do not contain this term,
can be written down through
the {\em second} derivative of the scalar correlator
\cite{Becchi}. It leads to the following
Euclidean function 
\begin{equation}
\overline{D}_S(Q^2)=2Q^2\int_0^\infty
\frac{sR_S(s)}{(s+Q^2)^3}ds~~~.
\label{Db}
\end{equation}
Note, however, that its  perturbative expansion differs from the one,
which corresponds
to  the  Euclidean part of perturbative series for
$\Gamma(H^0\rightarrow \overline{q}q)$, generated by
the ill-defined in non-perturbative sector
expression of Eq.(\ref{DS}).
Moreover, the application  of the  ``approximate'' dispersion relation from
Eq.(\ref{DS})
fixes  the kinematic $\pi^2$-contributions to
the coefficients of the perturbative   series for
$\Gamma(H^0\rightarrow \overline{q}q)$
both in the expanded \cite{Chetyrkin:1997wm} and summed up
\cite{Broadhurst:2000yc}, \cite{Bakulev:2006ex}  forms. Note, that
the idea of the  summation of   $\pi^2$-terms at lowest order of  QCD
was proposed and used
over thirty five  years ago in the works of Refs.
\cite{Radyushkin:1982kg}, \cite{Krasnikov:1982fx}, \cite{Gorishnii:1983cu}.
\section{Dispersion relations for the pole and \\ running heavy quark masses}
Consider now
the following  ``approximate'' dispersion
model of Ref.\cite{Chetyrkin:1997wm}   for the heavy quark pole  masses
\begin{equation}
\label{model}
M_{(N_L+1)}=\frac{1}{2\pi i}\int_{-\overline{m}_{(N_L+1)}
(\overline{m}_{(N_L+1)}^2)-i\varepsilon}^{-\overline{m}_{(N_L+1)}
(\overline{m}_{(N_L+1)}^2)+i\varepsilon}ds^\prime
   \int_0^\infty \frac{T(s)}{(s + s^\prime)^2}ds
\end{equation}
with the spectral density defined as
$T(s)=\overline{m}_{(N_L+1)}(s)\sum_{n=0}^{4}t_n^{M}a_s^n(s)$~.
It  can be obtained  from the dispersion-type expression
for the  {\bf Euclidean} series
\begin{equation}
F(Q^2)=\overline{m}_{(N_L+1)}(Q^2)\sum_{n=0}^{4} f_n^{E}(N_L)a_s^n(Q^2)=Q^2
\int_0^\infty \frac{T(s)}{(s + Q^2)^2}ds
\end{equation}
where $\overline{m}_{(N_L+1)}(Q^2)$ and $a_s(Q^2)$
are the heavy quark masses and the  QCD  coupling constant which are 
``running''   in the ${\bf Euclidean}$ region. The application
of Eq.(\ref{model}) allows one to fix the relations between coefficients
$f_n^{E}(N_L)$ and $t_n^{M}(N_L)$
of the perturbative series in the
time-like and space-like regions as
$f_0^{E}=t_0^{M}$~,$f_1^{E}=t_1^{M}$~,
$f_2^{E}(N_L)=t_2^{M}(N_L)+ e_2(N_L)$,
$f_3^{E}(N_L)=t_3^{M}(N_L)+e_3(N_L)$, $f_4^{E}(N_L)=t_4^{M}(N_L)+e_4(N_L)$.
The kinematic $\pi^2$-terms enter the  derived in
Ref.\cite{Chetyrkin:1997wm} explicit expressions for the
$e_i(N_L)$-contributions, namely ~~~
\begin{eqnarray}
\label{e2}
e_2(N_L)&=&\frac{\pi^2}{6}t_0^{M}\gamma_0(\beta_0+\gamma_0) \\ \nonumber
&=& 5.89435-0.274156N_L \\
\label{e3}
e_3(N_L)&=& \frac{\pi^2}{3}\left\{t_1^{M}(\beta_0+\gamma_0)
\left(\beta_0+\frac{\gamma_0}{2}\right)
+t_0^{M}
\left[\frac{\beta_1\gamma_0}{2}+\gamma_1(\beta_0+\gamma_0)\right]\right\} \\
&=& 105.622-10.0448N_L+0.198001N_L^2 \\
\label{e4}
e_4(N_L)&=&\pi^2 \bigg\{ t_2^{M}(\beta_0+\frac{\gamma_0}{2})+t_1^{M}\bigg[\frac
{\beta_1}{2}
(\frac{5}{3}\beta_0+\gamma_0)+\frac{\gamma_1}{3}(2\beta_0+
\gamma_0)\bigg]
  \\ \nonumber
&&  +t_0^{M}\bigg[\frac{\beta_2\gamma_0}{6}+\frac{\gamma_1}{3}\bigg(\beta_1+\frac{\gamma_1}{2}\bigg)+\gamma_2
\bigg(\frac{\beta_0}{2}+\frac{\gamma_0}{3}\bigg)\bigg]\bigg\} \\
\nonumber
&&+\frac{7\pi^4}{60}t_0^{M}\gamma_0(\beta_0+\gamma_0)(\beta_0
+\frac{\gamma_0}{2})(\beta_0+\frac{\gamma_0}{3}) \\ \nonumber
&=& 2272.02-403.951N_L+20.6768N_L^2-0.315898N_L^3
\end{eqnarray}
Their   $N_L$-dependence result  from $N_L$-dependence of the coefficients
 $\beta_i(N_L)$ with $i\geq 0$ in Eq.(\ref{beta}), $\gamma_i(N_L)$ with
$i\geq 1$ in  Eq.(\ref{gamma}) and $t_2^{M}$ in Eq.(\ref{massrel}),
which has the following numerical form \cite{Melnikov:2000qh}
\begin{equation}
t_2^{M}=13.44396-1.041367 N_L
\end{equation}
and  comes  from the  analytical expression  of Ref.\cite{Gray:1990yh},
confirmed by the independent calculations of Ref.\cite{Fleischer:1998dw}.
Notice, that the results of Refs.\cite{Gray:1990yh},
\cite{Fleischer:1998dw} contain the {\bf explicit dependence} from
$\zeta_2=\pi^2/6$-terms. The discussions presented above
clarify that the part of these  $\pi^2$-terms, explicitly visible in
the formulae of Refs.\cite{Gray:1990yh},\cite{Fleischer:1998dw},
appear from  the analytical continuation effect of Eq.(\ref{e2}).
This our claim can be  generalised to the level
of $t_3^{M}$-corrections, evaluated  analytically in Ref.\cite{Melnikov:2000qh}
and semi-analytically in Ref.\cite{Chetyrkin:1999qi}. In this case
kinematic $\pi^2$-contributions are determined by  Eq.(\ref{e3}).
The coefficients of the relation between 
heavy quark Euclidean masses, defined in the   MOM on-shell,   and
$\overline{\rm{MS}}$-scheme masses, contain only remaining 
transcendental terms,  typical to the on-shell scheme calculations.
\section{Estimates of $\alpha_s^4$ corrections}
We consider now two  perturbative series, namely the one 
of  Eq.(\ref{massrel}) and the related to it relation 
\begin{equation}
\label{Mv}
M_{(N_L+1)}=\overline{m}_{(N_L+1)}(M_{(N_L+1)}^2)\sum_{n=0}^{4}
v_n^{M}(N_L) a_s^n(M^2_{(N_L+1)})~~~.
\end{equation}
Keeping in mind that for $0\leq n \leq 3$  the values of the explicit
dependence from $N_L$ of the coefficients
$t_n^{M}(N_L)$ and $v_n^{M}(N_L)$ is already known
\cite{Melnikov:2000qh}, \cite{Chetyrkin:2000yt}, we will study the problem
of estimates of the $\alpha_s^3$ and   $\alpha_s^4$ coefficients, 
using the effective-charges (ECH)  inspired approach, developed and used  in 
Refs.\cite{Kataev:1995vh}, \cite{Chetyrkin:1997wm}{\footnote{ The method of ECH was proposed and 
developed in Refs.\cite{Grunberg:1980ja},\cite{Grunberg:1982fw} and 
independently in Ref.\cite{Krasnikov:1981rp} 
(see also Ref. \cite{Kataev:1981gr}).}. 
It is known  that the applications
of this approach in the Euclidean region  at the level of $\alpha_s^3$
and $\alpha_s^4$ corrections  give correct correct in signs  and in
order of magnitude estimates of the perturbative contributions to the
number of physical quantities  (see e.g. \cite{Kataev:1995vh},
\cite{Chetyrkin:1997wm},  \cite{Baikov:2005rw}, \cite{Baikov:2008jh}).
As to the application
of this procedure to the  Minkowskian quantities, two ways are possible.
The first,  prescribes to apply the
procedure of estimates  in the Euclidean region and {\bf add explicitly
calculable} kinematic $\pi^2$-terms afterwards. Within the second way
one may   use the procedure of estimates in the Minkowski region
directly. It should be noted, that both ways are leading 
to reasonable
predictions of  signs and numerical values of perturbative 
series for physical quantities.  Moreover, in the case
of {\bf direct} application of this approach in the Minkowski
region, the order $\alpha_s^4$ estimates are sometimes  even closer to the
results of the explicit calculations (see e.g. Ref. \cite{Baikov:2005rw}).
However, in the latter case the estimates do not reproduce
the known  values of the analytical continuation effects, similar to the 
ones of Eq.(\ref{e3}) and Eq.(\ref{e4}).  Note, that
their precise knowledge is important  
for applying different approaches
of  resummations of   these contributions 
(see e.g. \cite{Pivovarov:1991rh}-\cite{Bakulev:2008td},
\cite{Broadhurst:2000yc}, \cite{Bakulev:2006ex}).
Following two ways mentioned above  we  first estimate the values
of  $t_3^{M}(N_L)$ coefficients  and compare them with the 
results for $t_3^{exact}(N_L)$ obtained  in  Refs.\cite{Melnikov:2000qh},
\cite{Chetyrkin:1999qi}. 
Satisfied by this  comparison we are going one 
step further and estimate $t_4^{M}(N_L)$-coefficients,  
taking into account the numerical expressions  for $t_3^{exact}(N_L)$.  
The concrete numbers  are presented in  Table 1.
\begin{table}[h!]
\caption{The estimates for $t_3^{M}(N_L)$, $t_4^{M}(N_L)$.}
\begin{tabular}{|c|c|c|c|c|c|}
\hline
$N_L$ \rule{0pt}{15pt} & $t_3^{exact}$ & $t_3^{ECH}$ &   $t_3^{ECH \, direct}$ &  $t_4^{ECH}$ &  $t_4^{ECH \, direct}$   \\
\hline
\hline
5 \rule{0pt}{15pt} & 73.6366 & 58.0645 &  48.4906   & 719.339 & 710.016 \\
\hline
4  \rule{0pt}{15pt} & 94.4175 & 100.74 &  78.243   & 986.097 & 1045.5  \\
\hline
3  \rule{0pt}{15pt} & 116.504 & 147.303 &111.315   &    1281.05 & 1438.75     \\
\hline
\end{tabular}
\label{tab:t3}
\end{table}
One can see, that the estimates obtained give correct signs and order of 
magnitude estimates for the values of  $t_3^{M}(N_L)$-terms. 
Thus, one may hope that  the estimates for  $t_4^{M}(N_L)$
are not far from reality.
We present  now   concrete numbers for the coefficients of the  series of
Eq.(\ref{massrel}), where for the   $\alpha_s^4$-coefficients we use the 
estimates  
$t_4^{ECH}(N_L)$ from Table 1: 
\begin{equation}
M_c\approx\overline{m}_{c}(\overline{m}_{c}^2)
\bigg[1+\frac{4}{3}a_s(\overline{m}_{c}^2)+
10.3a_s^2(\overline{m}_{c}^2)+116.5a_s^3(\overline{m}_{c}^2)
+1281a_s^4(\overline{m}_{c}^2)\bigg] 
\end{equation}
\begin{equation}
M_b\approx\overline{m}_{b}(\overline{m}_{b}^2)\bigg[1
+\frac{4}{3}a_s(\overline{m}_{b}^2)+
9.28a_s^2(\overline{m}_{b}^2)+94.4a_s^3(\overline{m}_{b}^2)
+986a_s^4(\overline{m}_{b}^2)\bigg] 
\end{equation}
\begin{equation}
M_t\approx \overline{m}_{t}(\overline{m}_{t}^2)\bigg[1+\frac{4}{3}
a_s(\overline{m}_{t}^2)+
8.24a_s^2(\overline{m}_{t}^2)+73.6a_s^3(\overline{m}_{t}^2)+
719a_s^4(\overline{m}_{t}^2)\bigg] 
\end{equation}
The similar relations for Eq.(\ref{Mv}) with  on-shell normalizations of running 
parameters read   
\begin{equation}
\label{Mc}
M_c \approx \overline{m}_{c}(M_c^2)
\bigg[1+\frac{4}{3}a_s(M_c^2)+13a_s^2(M_c^2)+156a_s^3(M_c^2)+1853a_s^4(M_c^2)\bigg] 
\end{equation}
\begin{equation}
\label{Mb}
M_b \approx \overline{m}_{b}(M_b^2)\bigg
[1+\frac{4}{3}a_s(M_b^2)+12a_s^2(M_b^2)+131a_s^3(M_b^2)+1460a_s^4(M_b^2)\bigg] 
\end{equation}
\begin{equation}
\label{Mt}
M_t \approx \overline{m}_{t}(M_t^2)\bigg[1+\frac{4}{3}a_s(M_t^2)+
11a_s^2(M_t^2)+107a_s^3(M_t^2)+1101a_s^4(M_t^2)\bigg]
\end{equation}
The results presented in Eq.(\ref{Mb})  give the 
following ratios of the squares of running and
pole $b$-quark masses
\begin{equation}
 \label{mb}
 \frac{\overline{m}_b^2(M_b^2)}
      {M_b^2}
  = 1-\frac{8}{3} a_s(M_b^2)
     -18.5559 a_s^2(M_b^2)
     -175.797 a_s^3(M_b^2)
     -1684   a_s^4(M_b^2)\,
\end{equation}
where the last term is fixed by the result of application of  
the ECH-motivated approach with adding    
kinematic $\pi^2$-contributions at the final step.
In the case when the Euclidean and kinematic $\pi^2$ corrections 
are summed up at the intermediate steps,   the last coefficient in Eq.(\ref{mb}) should be changed 
from {\bf -1684} to  {\bf -1835}.
Note, that in the process of analysing   the uncertainties of QCD 
predictions for $\Gamma(H^0\rightarrow
b\overline{b})$, preformed in the work of Ref.\cite{Kataev:2009ns}, 
we used slightly lower
estimate, namely {\bf -1892}. The difference is explained
in part by smaller number of significant digits taken into account  
in the values of 
coefficients, which
enter in the procedure of corresponding estimates. However, this difference 
between the values of estimated order $O(\alpha_s^4)$ contributions  
are not so numerically important.
Other possible physical applications, like the comparison with the 
renormalon-based  analysis  
of asymptotic behaviour of perturbative series in  
Eq.(\ref{Mc})-Eq.(\ref{Mt}) (for the related
theoretical discussions one can see  
Refs. \cite{Beneke:1994sw}-\cite{Hoang:2008yj}) are beyond the scope of this 
study.
\section{Acknowledgements}
The results described above were presented by one of us (ALK) at the International Conference ``Problems
of Theoretical and Mathematical Physics'', dedicated to the 100th anniversary 
of the birth of N.N. Bogolyubov, Dubna, August 23-27, 2009,
and  at the 
9th International Symposium on Radiative Corrections: Applications of Quantum Field Theory to Phenomenology 
(RADCOR-2009), October 25-30, 2009,
 Ascona, Switzerland.
It is the pleasure to thank the Organizers of these scientific events for 
hospitality and partial financial support. We are grateful to A.P. Bakulev ,  
M. Y. Kalmykov and S.V. Mikhailov for their interest to our
work, and to J. Bluemlein for the questions, which stimulated us to describe in more detail
the origin of the appearance of the $\pi^2$-analytical continuation
effects in the relations of heavy quark masses, defined in the Minkowski region.
We also wish to thank N.V. Krasnikov for discussions of the topics concerning   
different applications of  the K\"allen-Lehman representations. 
The work is done as the research,
planned  within the
program of the Grants  RFBR  08-01-00686, 08-02-01184 and
President of RF NS-1616.2008.2 and NS-378.2008.2.

\end{document}